\documentclass[sigconf,screen]{acmart}
\usepackage{subcaption}
\usepackage{balance}
\usepackage[shortlabels]{enumitem}
\usepackage{multirow}
\setcopyright{acmcopyright}
\acmPrice{15.00}
\acmDOI{10.1145/3368089.3417943}
\acmYear{2020}
\copyrightyear{2020}
\acmSubmissionID{fse20demo-p46-p}
\acmISBN{978-1-4503-7043-1/20/11}
\acmConference[ESEC/FSE '20]{Proceedings of the 28th ACM Joint European Software Engineering Conference and Symposium on the Foundations of Software Engineering}{November 8--13, 2020}{Virtual Event, USA}
\acmBooktitle{Proceedings of the 28th ACM Joint European Software Engineering Conference and Symposium on the Foundations of Software Engineering (ESEC/FSE '20), November 8--13, 2020, Virtual Event, USA}

\begin{document}

\title[BugsInPy: A Database of Existing Bugs in  Python Programs]{BugsInPy: A Database of Existing Bugs in  Python Programs to Enable Controlled Testing and Debugging Studies}
\author{Ratnadira Widyasari}
% \affiliation{
%   \institution{Singapore Management University, Singapore}
% }
%\email{ratnadiraw.2020@smu.edu.sg}
\author{Sheng Qin Sim}
% \affiliation{
%   \institution{Singapore Management University, Singapore}
% }
\author{Camellia Lok}
% \affiliation{
%   \institution{Singapore Management University, Singapore}
% }
\author{Haodi Qi}
\affiliation{%
  \institution{Singapore Management University, Singapore}
}
\author{Jack Phan}
%\email{jack.phan.2018@smu.edu.sg}
% \affiliation{
%   \institution{Singapore Management University, Singapore}
% }
\author{Qijin Tay}
% \affiliation{
%   \institution{Singapore Management University, Singapore}
% }
\author{Constance Tan}
% \affiliation{
%   \institution{Singapore Management University, Singapore}
% }
\author{Fiona Wee}
\affiliation{
  \institution{Singapore Management University, Singapore}
}
\author{Jodie Ethelda Tan}
%\email{jodie.tan.2018@smu.edu.sg}
% \affiliation{
%   \institution{Singapore Management University, Singapore}
% }
\author{Yuheng Yieh}
% \affiliation{
%   \institution{Singapore Management University, Singapore}
% }
\author{Brian Goh}
% \affiliation{
%   \institution{Singapore Management University, Singapore}
% }
\author{Ferdian Thung}
\affiliation{
  \institution{Singapore Management University, Singapore}
}
\author{Hong Jin Kang}
%\email{hjkang.2018@smu.edu.sg}
% \affiliation{
%   \institution{Singapore Management University, Singapore}
% }
\author{Thong Hoang}
% \affiliation{
%   \institution{Singapore Management University, Singapore}
% }
\author{David Lo}
% \affiliation{
%   \institution{Singapore Management University, Singapore}
% }
\author{Eng Lieh Ouh}
\affiliation{%
  \institution{Singapore Management University, Singapore}
}

\begin{abstract}
The 2019 edition of Stack Overflow developer survey highlights that, for the first time, Python outperformed Java in terms of popularity. 
The gap between Python and Java further widened in the 2020 edition of the survey. 
Unfortunately, despite the rapid increase in Python's popularity, there are not many testing and debugging tools that are designed for Python. 
This is in stark contrast with the abundance of testing and debugging tools for Java. Thus, there is a need to push research on tools that can help Python developers.

One factor that contributed to the rapid growth of Java testing and debugging tools is the availability of benchmarks. 
A popular benchmark is the Defects4J benchmark; its initial version contained 357 real bugs from 5 real-world Java programs. 
Each bug comes with a test suite that can expose the bug. 
Defects4J has been used by hundreds of testing and debugging studies and has helped to push the frontier of research in these directions.

In this project, inspired by Defects4J, we create another benchmark database and tool that contain 493 real bugs from 17 real-world Python programs. We hope our benchmark can help catalyze future work on testing and debugging tools that work on Python programs.
\end{abstract}

\maketitle
\renewcommand{\shortauthors}{Widyasari, et al.}

\section{Introduction}
\label{sec:intro}

Python is among one of the most popular programming languages in the world today\footnote{\url{https://www.tiobe.com/tiobe-index/}}$^,$\footnote{\url{https://insights.stackoverflow.com/survey/2020}}.
Understanding the bugs and faults in large software repositories built in Python is therefore important. 
Python has been largely overlooked in the software engineering research community and disproportionately little effort has been given to studies on software projects primarily written in Python. 
Python has features, such as duck typing and common use of heterogeneous collections, that distinguish it from other popular languages. 
It is used in diverse domains, spanning the most popular machine learning libraries and popular web frameworks.
As a result, the characteristics of bugs that occur in Python projects are likely to differ from bugs in other programming languages.  
This highlights the need for more research on projects using the Python programming language.

A collection of known bugs is required to evaluate automated testing and debugging solutions.
To support reproducible research, it is crucial that studies are tested empirically on similar,  publicly-available data.
In the absence of a curated dataset, researchers must collect bugs that are reproducible from open-source repositories, which is a highly time-consuming process.

In this work, we attempt to reduce the barrier of entry for research and development of testing and debugging tools targeting Python programs.
We propose BugsInPy, inspired by Defects4J~\cite{just2014defects4j} which was originally proposed to support software testing research for Java programs.
After its release, Defects4J has been used by hundreds of studies, primarily as an evaluation benchmark. This includes studies on software testing~\cite{just2014mutants,ma2015grt,lu2016does}, fault localization~\cite{b2016learning,sohn2017fluccs,xia2016automated} and automated program repair~\cite{le2016history,martinez2017automatic,xiong2017precise} targeting Java programs.
Its popularity shows that many researchers find it useful. 
This is, in part, due to the high quality of the bugs in Defects4J.
Firstly, the bugs in Defects4J come from real-world projects. 
Secondly, other than providing the buggy programs, Defects4J ensures that the bugs are \textit{reproducible}, and each is accompanied by a failing test case that passes once the bug is fixed. 
Thirdly, the bugs are \textit{isolated}, and the code changes that fix the bugs do not contain irrelevant changes.
Finally, apart from the quality of the dataset, Defects4J makes it easy to retrieve each project at its buggy revision as well as obtain the corresponding test suite that exposes the bug.
We construct BugsInPy taking care to ensure that it has the same quality as Defects4J.

BugsInPy currently has 493 bugs from 17 real-world Python projects. 
These projects were selected as they represent the diverse domains (machine learning, developer tools, scientific computing, web frameworks, etc) that Python is used for. 
These projects are Python open-source projects on GitHub, each with more than 10,000 stars.
Constructing and manually validating the bugs and test cases for this dataset required significant effort, and took an estimated 831 man-hours.
Another key feature of BugsInPy is its extensibility. 
Much like Defects4J, BugsInPy is an extensible framework that simplifies access to revisions of a project, before- and after- a bug fixing commit. 
Adding a new bug into BugsInPy is simple and requires only some configurations in the form of records of commands to setup the project and run the test cases.
A guide on how to add a new bug is available in the BugsInPy repository.%\footnote{\url{https://github.com/soarsmu/BugsInPy}}

\begin{figure}[t]
    \begin{center}
        \includegraphics[width=0.7\linewidth]{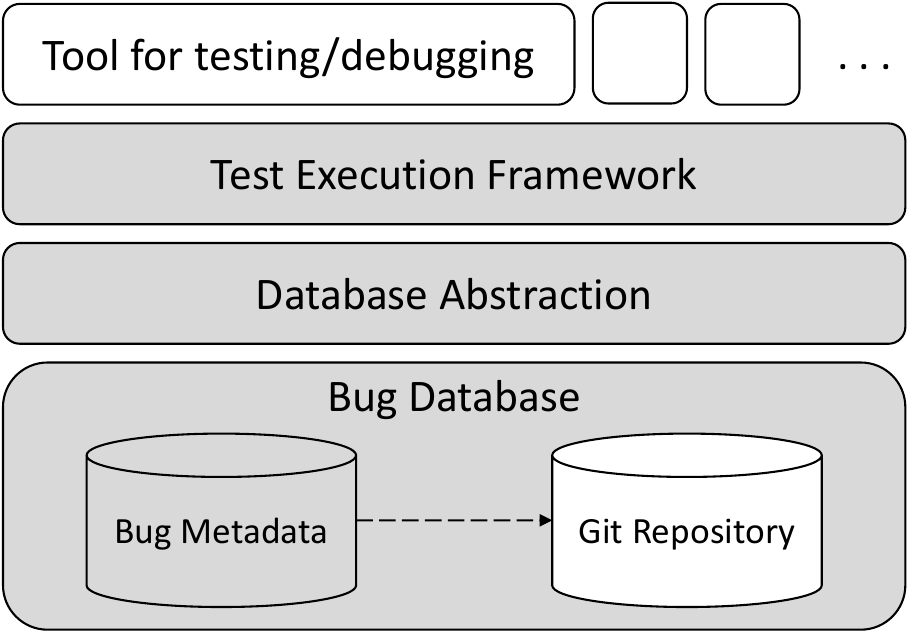}
        \caption{Architecture of BugsInPy}
        \label{fig:architecture}
    \end{center}
    
\vspace{-0.4cm}
\end{figure}

BugsInPy's architecture is similar to Defects4J, as shown in Figure~\ref{fig:architecture}. It has three main components (highlighted in gray): a bug database, a database abstraction layer, and a test execution framework. The bug database contains the collected bug metadata with links to the original Git repositories. The database abstraction layer allows access to bugs without the knowledge on how the bug data is stored. It abstracts details on how to checkout and build faulty or fixed source code versions. The test execution framework allows execution of tools for testing/debugging on the collected bug data. It currently supports test execution, test input generation, mutation analysis, and code coverage analysis. 

We make the following contributions in this work:
\begin{itemize}[nosep,leftmargin=*]
    \item BugsInPy contains a hand-curated dataset of real-world bugs in large, non-trivial Python projects. These bugs are reproducible and isolated.
    \item BugsInPy makes it easy to retrieve the buggy versions of a project and run the test cases that reveal the bugs.
    \item BugsInPy makes it easy to extend the dataset. The projects we study are actively developed. As they continue to evolve, the new bug fixes can be added into BugsInPy.
    \item BugsInPy makes it easy to run test cases, compute code coverage, perform mutation analysis, and generate new test inputs via its integration with existing tools.
    
\end{itemize}

The remainder of this paper is structured as follows. Section~\ref{sec:method} describes how we obtained the bug data for BugsInPy.  Sections~\ref{sec:db},~\ref{sec:dblayer}, and~\ref{sec:test_framework} describe the bug database, the database abstraction layer, and the test execution framework. Section~\ref{sec:threat} describes threats to validity. Some related work are presented in Section~\ref{sec:related_work}. Finally, we conclude and mention some future work in Section~\ref{sec:conclusion}.

\section{Detecting bugs from version control history}
\label{sec:method}

In this section, we briefly describe the framework used to construct BugsInPy's bug database. We also highlight challenges in collecting and reproducing real bugs from version control history and how we address these challenges. Our goal is to obtain bugs fixed by developers. For each bug in our database, we wish to identify a faulty and a developer-fixed source code version. Specifically, each bug in BugsInPy should fulfill the following requirements:
\begin{enumerate}[nosep,leftmargin=*]
    \item {\em The bug is in source code}. We include only bug fixes involving changes in source code and exclude those that change configurations, build scripts, documentation, and test cases.
    \item {\em The bug is reproducible}. At least one of the test cases from the fixed version should fail on the faulty version.
    \item {\em The bug is isolated}. The faulty and fixed versions differ only by code changes required to fix the bug and no other unrelated changes are involved (e.g., refactoring or feature addition).
\end{enumerate}
We populate BugsInPy with real bugs recorded in  version control systems by employing several  strategies to fulfill the above requirements. 

%\subsection{Identify Real Bugs}
\vspace{0.2cm}\noindent\textbf{Identify Real Bugs.}
When collecting bugs, we investigate commits that modify or add test files. Such commits are good starting points in our search of bugs that are reproducible by a test case. We heuristically identify test files as files that contain ``test'' in their names and import testing library such as unittest\footnote{\url{https://docs.python.org/3/library/unittest.html}} or pytest\footnote{\url{https://docs.pytest.org/en/stable/}}. For each commit, we need to identify whether it fixes a bug. To identify whether a commit is a bug fix, we manually look at the commit message, the source code, and any linked information such as GitHub issues to understand the intention of the changes introduced by the commit. The link to a Github issue is optional since not all projects links its bug-fixing commit to a GitHub issue (i.e., a bug report). One of the challenges in identifying bug fixes that satisfy requirement (1) is that developers may also label fixes on build scripts, configuration files, test cases, and documentations as bug fixes. These labels could appear in the commit message or in the corresponding issue tracking system. To exclude these cases, we only look at changes on ``*.py'' files (i.e., Python source code files) that are not test files. Moreover, to further ensure that we identify real bug fixes that satisfy requirement (1), at least two authors investigate the commits independently and we take only the commits that they agree on as qualifying bug-fixing commits. 
In this step, we identified 796 commits initially, and 66 commits were omitted as the authors did not agree that they qualified based on our criteria.

%\subsection{Reproduce Real Bugs}
\vspace{0.2cm}\noindent\textbf{Reproduce Real Bugs.}
To satisfy requirement (2), a bug fixing commit should contain at least a test case that exposes the bug. We identify these test cases by running them on both the faulty and fixed source code versions. These test cases should fail on the faulty source code version and run successfully on the fixed source code version. We identify these test cases as the ones that trigger the bug. We exclude bug fixing commits that do not have such test cases. 

%\subsection{Isolate Real Bugs}
\vspace{0.2cm}\noindent\textbf{Isolate Real Bugs.}
A bug fixing commit may contain changes other than bug fixes, e.g.,  refactoring and feature addition. In such a case, the bug fixing commit is not isolated and thus does not satisfy requirement (3). We include only bug fixing commits that are isolated, as otherwise the failed test cases may fail because of other reasons such as non-existence of a new feature in the faulty source code version. To identify the isolated bug-fixing commits, two authors manually check the commits and label whether the commits also contain refactoring and feature addition. The commits are considered to be isolated if both the authors reach the same conclusion independently. Commits that are not selected as isolated commits are not necessarily harder to fix.
These commits are not selected because of the lack of consensus between two authors investigating the commits about whether they contain unrelated changes, such as refactoring.
Of the 730 commits collected in the previous step, 235 of them were omitted in this step as the 
two 
authors did not agree 
if the commits do not contain unrelated changes.
As an alternative, it is possible to manually ``clean'' such {\em tangled} commits, e.g., by removing refactoring and feature addition from them. However, we choose not to do so as we want all buggy and fixed versions in our database to be real (i.e., they appear in the version control system of a real project).

\section{Database of Real Python Bugs}
\label{sec:db}
Our BugsInPy database contains 493 real Python bugs from 17 open-source projects. We selected Python 3 projects from GitHub with a high number of stars ($>$10K) and available in PyPI\footnote{\url{https://pypi.org/}}, a repository of software for the Python programming language. For each project's repository, we only investigated commits from its master branch. 
Table~\ref{tab:dataset} shows the statistics of the projects and number of real bugs available in BugsInPy. KLoC is counted based on the version downloaded on 19 June 2020, as reported by SLOCCount\footnote{\url{https://dwheeler.com/sloccount/}}.

BugsInPy provides the following artifacts and metadata for each bug in each project:
\begin{itemize}[nosep,leftmargin=*]
    \item {{\em Revisions in the project's version control system.} Our bug database has its own bug id for each bug in the project. We maintain the mapping of this bug id to the Git revision hash in its original GitHub repository.}
    \item {\em Patch of isolated bug.} Our bug database provides the original patch that fixes the bug. The patch is taken from the {\em diff} of source code files (i.e., excluding test files) between the faulty and fixed versions.
    \item {\em Tests that expose the bug.} Our bug database has a list of test cases that expose the bug. 
\end{itemize}

\setlength{\tabcolsep}{1.2pt}
\begin{table}[!t]
\caption{Projects and number of real bugs available in the initial version of BugsInPy (as of 19 June 2020) }
\label{tab:dataset}
\centering
\vspace{-0.4cm}
\small
\begin{tabular}{|l|l|l|l|l|l|}
\hline
\textbf{Project} & \textbf{Bugs} & \textbf{LoC} & \textbf{Test LoC} & \textbf{\# Tests} & \textbf{\# Stars} \\
% \hline
\hline
ansible/ansible~\cite{ansible_git} & 18 & 207.3K                          & 128.8K                               &   20,434          & 43.6K                            \\
cookiecutter/cookiecutter~\cite{cookiecutter_git} & 4 & 4.7K & 3.4K & 300  & 12.2K \\
cool-RR/PySnooper~\cite{rachum2019pysnooper}                  & 3                               & 4.3K                            & 3.6K                                 &      73                              & 13.5K                            \\
explosion/spaCy~\cite{Honnibal_spaCy_Industrial-strength_Natural_2020}                    & 10                               & 102K                            & 13K                                  &  1,732                            & 16.6K                            \\
huge-success/sanic~\cite{sanic_git}                 & 5                               & 14.1K                           & 8.1K                                 &  643                  & 13.9K                            \\
jakubroztocil/httpie~\cite{httpie_git}               & 5                               & 5.6K                            & 2.2K                         &   309     & 47K                              \\
keras-team/keras~\cite{chollet2015keras}                   & 45                              & 48.2K                           & 17.9K                                & 841      & 48.6K                            \\
matplotlib/matplotlib~\cite{Hunter_Matplotlib_A_2D_2007} & 30 & 213.2K & 23.2K & 7,498 & 11.6K \\
nvbn/thefuck~\cite{thefuck_git}                       & 32                               & 11.4K                           & 6.9K                                 &  1,741                          & 53.9K                            \\
pandas-dev/pandas~\cite{The_pandas_development_team_pandas-dev_pandas_Pandas}                  & 169                             & 292.2K                          & 196.7K                               &  70,333                          & 25.4K                            \\
psf/black~\cite{Langa_Black_The_uncompromising}                          & 15                              & 96K                             & 5.8K                                 &    142                  & 16.4K                            \\
scrapy/scrapy~\cite{scrapy_git}                      & 40                              & 30.7K                           & 18.6K                                &  2,381                     & 37.4K                            \\
spotify/luigi~\cite{luigi_git}                      & 33                               & 41.5K                           & 20.7K                                &   1,718                       & 13.4K                            \\
tiangolo/fastapi~\cite{Ramirez_FastAPI}                   & 16                              & 25.3K                           & 16.7K                                & 842                          & 15.3K                            \\
tornadoweb/tornado~\cite{tornado_git}                 & 16                              & 27.7K                           & 12.9K                                &   1,160                & 19.2K                            \\
tqdm/tqdm~\cite{da2019tqdm}                          & 9                               & 4.8K                            & 2.3K                                 & 88                     & 14.9K                            \\
ytdl-org/youtube-dl~\cite{youtube-dl_git}                & 43                              & 124.5K                          & 5.2K                                 & 2,367        & 67.3K                            \\
\hline
\textbf{Total}                              & 493                             & 1253.5K                         & 486K                 &     112,602                                         & 470.2K\\
\hline
\end{tabular}
\vspace{-0.2cm}

\end{table}

\section{Database Abstraction Layer}
\label{sec:dblayer}
BugsInPy abstracts away access to bug artifacts via a database abstraction layer. This abstraction layer allows users to access the faulty and fixed source code versions, compile the source code, and test the source code without the knowledge of the underlying commands and technologies. 

The database abstraction layer provides the following components to access the bug artifacts:
\begin{itemize}[nosep,leftmargin=*]
    \item {\em Abstraction of source code access.} This component provides an interface to checkout the faulty and the fixed source code versions without the knowledge of the original repository location and Git revision hash.
    \item {\em Abstraction of build systems.} This component provides an interface to compile the source code without the knowledge of commands to run and dependencies to install. It also provides an interface to run test cases without knowing the underlying test automation framework. 
\end{itemize}

To abstract away source code access, BugsInPy assigns a unique id to each bug in a project. Internally, the unique id is linked to the Git revision hash in the original project repository. When a user requests for a source code version (i.e., either faulty or fixed), BugsInPy finds the Git revision hash that is linked to the id and checkout the source code from the original project repository that corresponds to the Git revision hash. 

To abstract away build systems, we manually investigate the project and learn how to build it. The learning process involves reading the documentation (i.e., in the project readme or website) and potentially looking through the source code. We record how to build each project and automate the process, thus removing the need for users to manually configure each project themselves. 

The build process consists of compiling the project and running test cases.
To compile a project, we install the required dependencies listed in the {\tt requirements.txt} (i.e., the file listing the versions of project dependencies). 
We may also run {\tt setup.py} (i.e., a standard Python setup script) with differing arguments depending on the project. 
To run test cases, we first need to figure out the test automation framework used by the project. 
There are two frameworks used by projects in the initial version of BugsInPy: unittest\footnote{\url{https://docs.python.org/3/library/unittest.html}} and pytest\footnote{\url{https://docs.pytest.org/en/stable/}}. The commands to run test cases depend on which framework is used by a project, which is abstracted away by BugsInPy.
\section{Test Execution Framework}
\label{sec:test_framework}

BugsInPy provides a test execution framework to support common tasks in testing and debugging. 
The purpose of this framework is to minimize effort to run these common tasks, which include test set selection, test input generation, mutation analysis, and code coverage computation. 
To support these tasks, BugsInPy integrates existing tools into its test execution framework.

The test execution framework provides the following components for testing and debugging:
\begin{itemize}[nosep,leftmargin=*]
    \item {\em Test Set Selection.} This component provides an interface to select a set of test cases for execution. 
    It allows users to run a single test case, all test cases, or any subset of test cases. The selected test cases can be run in any faulty or fixed source code version.
    \item {\em Test Input Generation.} This component supports the generation of new test inputs via fuzzing. Test inputs can be generated for any faulty or fixed source code version. BugsInPy employs PythonFuzz\footnote{\url{https://github.com/fuzzitdev/pythonfuzz}}, a coverage-guided fuzzer as the test input generator.
    \item {\em Mutation Analysis.} This component supports mutation analysis for any test case on any faulty or fixed source code version. BugsInPy employs MutPy\footnote{\url{https://github.com/mutpy/mutpy}} as the mutation testing tool.
    \item {\em Code Coverage Computation.} This component supports the measurement of code coverage for any set of test cases on any faulty or fixed source code version. BugsInPy employs coverage.py\footnote{\url{https://coverage.readthedocs.io/en/coverage-5.1/}} as the code coverage tool. 
\end{itemize}

The test execution framework runs on top of the database abstraction layer (see Section~\ref{sec:dblayer}). 
Therefore, it can access any faulty or fixed source code versions and any test cases via the database abstraction layer. 
It further provides an abstraction for running external testing tools and managing their generated data.
\section{Threats to Validity}
\label{sec:threat}
To ensure the quality of our bug data, we manually curate the bugs in BugsInPy. Yet, despite our best effort, we may still mislabel the bug (i.e., include bugs that do not satisfy the three requirements in Section~\ref{sec:method}). 
To minimize the risk of mislabelling, we require two authors working  independently to agree and be confident on any labelling decision, either when deciding whether a commit is indeed a bug fix or when deciding whether a bug is isolated. 
If consensus is not reached, we discard the bug from our dataset. 
In other words, we only include bugs that we are highly confident about.

Any program may contain bugs, including the ones supporting our benchmark (e.g., the test execution framework). 
We have tried to ensure our programs are bug-free and have checked them multiple times. 
Yet, there may still be bugs that we did not encounter.
\section{Related Work}
\label{sec:related_work}

A popular bug database is the Software-artifact Infrastructure Repository (SIR)~\cite{do2005supporting} containing 81 bugs that appear in programs written in Java, C, C$++$, and C$\#$. However, only 35 bugs are real bugs and the remaining ones are obtained via mutation analysis. 
The Siemens benchmark suite~\cite{hutchins1994experiments} is another bug database. However, it only includes bugs for C programs and all the bugs are synthetic (i.e., they are artificially seeded into the programs). 

As described earlier,
Defects4J~\cite{just2014defects4j} is the closest related work, 
containing 357 real bugs from 5 real-world Java programs.
Another Java-focused bug dataset is Bugs.Jar~\cite{saha2018bugs}, which contains 1,158 bugs from 8 large and popular open-source Java projects.
Our work is inspired by Defects4J, and we strive to ensure that BugsInPy is of similar quality so it can follow Defects4J footsteps to be the first benchmark of its kind for Python.

Recently, Tomassi et al.~\cite{tomassi2019bugswarm} has proposed Bugswarm, which automatically mines failing and subsequently passing builds on Travis. 
This enables the collection of reproducible bugs in open-source projects.
While Bugswarm contains bugs in Python projects, it has several limitations, as pointed out by Durieux and Abreu~\cite{durieux2019critical}.
One limitation was the high cost of downloading the many Docker containers, one for each bug.
BugsInPy avoids this cost, by providing only one Docker container, which is available at \url{https://hub.docker.com/r/soarsmu/bugsinpy}.
Furthermore, while their approach finds many pairs of failing and passing builds, 
many of Bugswarm's bugs are duplicates of each other, contain only modifications to test cases, are due to compilation errors, or are not isolated. 
BugsInPy avoids all of these issues as its bug fixes are {\em manually curated} to ensure its quality. 

BugsJS~\cite{gyimesi2019bugsjs} was proposed recently to provide researchers with a benchmark of bugs in the JavaScript ecosystem. 
Similar to our work, BugsJS aims to fill the void of a good benchmark in its target programming language, providing 453 real bugs from 10 JavaScript programs.
Defexts~\cite{benton2019defexts} was proposed recently for Kotlin and Groovy, providing 225 Kotlin and 301 Groovy bugs.

QuixBugs~\cite{lin2017quixbugs} is a benchmark including small programs in Java and Python, it has bugs that can be fixed by changing a single line of code. These programs are not real software projects, but rather synthetically created programs of 17-48 lines of code. Moreover, the bugs are seeded into the programs.
In contrast, BugsInPy has 493 real bugs from 17 popular Python projects.

\section{Conclusion and Future Work}
\label{sec:conclusion}
% \balance
To conclude, we present BugsInPy, a framework to enable controlled studies requiring experiments on real bugs in Python projects
, such as work on testing and debugging.
The objective of this work is to support reproducible research on real-world Python projects.
BugsInPy is built to be extensible and currently comprises 493 bugs from 17 real-world projects, making it the largest Python bug dataset to date.
It is curated by hand to ensure that the bugs are reproducible and isolated.

In the future, we plan to add more projects and bugs to BugsInPy. 
Adding new projects and bugs into BugsInPy requires some manual effort. 
Fortunately, this is a one-time effort, after which the bugs can be reproduced easily. 
We also plan to integrate BugsInPy with more testing and debugging tools. 
We hope BugsInPy can stimulate the rapid growth of testing and debugging tools that target Python programs. 
BugsInPy is available at \url{https://github.com/soarsmu/BugsInPy}.

\section*{Acknowledgement}
This research is partially supported by the Lee Kuan Yew Fellowship awarded by Singapore Management University.
\balance
\bibliographystyle{ACM-Reference-Format}
\bibliography{bib}
\end{document}